# Pressure induced phase transitions in Sm-doped BiFeO$_3$ in the morphotropic phase boundary

A. Pakalniškis[a], R. Skaudžius[a], D.V. Zhaludkevich[b], S.I. Latushka[b], V. Sikolenko[c, d, e], A.V. Sysa[f, g], M. Silibin [f, g], K. Mažeika[h], D. Baltrūnas[h], G. Niaura[i], M. Talaikis[j], D.V. Karpinsky[b,k], A. Kareiva[a]

[a] *Institute of Chemistry, Vilnius University, Naugarduko 24, LT-03225 Vilnius, Lithuania*
[b] *Scientific-Practical Materials Research Centre of NAS of Belarus, 220072 Minsk, Belarus*
[c] *REC "Functional nanomaterials", Immanuel Kant Baltic University, 236041 Kaliningrad, Russia*
[d] *Joint Institute for Nuclear Research, 141980 Dubna, Russia*
[e] *Karlsruhe Institute of Technology, 76131 Karlsruhe, Germany*
[f] *National Research University of Electronic Technology "MIET", 124498 Zelenograd, Moscow, Russia*
[g] *Scientific-Manufacturing Complex "Technological Centre", 124498 Zelenograd, Moscow, Russia*
[h] *Center for Physical Sciences and Technology, Vilnius, LT-02300, Lithuania*
[i] *Institute of Chemical Physics, Faculty of Physics, Vilnius University, Sauletekio Ave. 3, LT-10257, Vilnius, Lithuania*
[j] *Department of Bioelectrochemistry and Biospectroscopy, Institute of Biochemistry, Life Sciences center, Vilnius University, Sauletekio Ave 7, LT-10257 Vilnius, Lithuania*
[k] *South Ural State University, av. Lenina, 76, 454080, Chelyabinsk, Russia*

*Corresponding author:*
A. Pakalniškis (*pakalniskis.andrius@chgf.vu.lt*);

## Abstract

Sm-doped BiFeO$_3$ compacted powders with composition across the morphotropic phase boundary region were prepared by sol-gel method. Crystal structure, morphology and magnetic state of the compounds were analyzed as a function of dopant concentration, temperature and external pressure using synchrotron and laboratory X-ray diffraction, electron microscopy, Raman and Mossbauer spectroscopy. Application of external pressure shifts the phase transition from the rhombohedral structure to the nonpolar orthorhombic structure towards lower concentration of the dopant content, wherein the amount of the anti-polar orthorhombic phase notably decreases. Raman and Mossbauer spectroscopy data provides additional information about the structural distortion on local scale level which testifies faster formation of nonpolar orthorhombic phase and associated modification in the magnetic state in the compounds subjected to high pressure. Temperature increase leads to the structural transition to the nonpolar orthorhombic phase regardless the structural state at room temperature; furthermore, application of external pressure decreases the phase transition temperature.

## 1. Introduction

The ever expanding need for materials in niche areas of application that require specific properties as well as stricter legislation on the use of hazardous chemicals has been a big driving force in material research [1–3]. That is especially the case for multiferroic compounds which exhibit at least two primary ferroic orders [4]. Single phase multiferroic compounds should simultaneously possess ferroelectric and ferromagnetic orders thus leading to magnetoelectric phenomenon and a possibility to control polarization by magnetic field and vice versa [5–7]. However, despite the number of performed investigations in this field there are only a few multiferroic compounds that have been discovered [8]. It should be noted that magnetic and electric orderings arise from contradicting phenomena [9]. For a strong magnetic ordering to exist partially filled electronic shells are needed while for ferroelectric ordering filled ones are required [9,10]. This problem was solved by the discovery of the fact that ferroelectricity can be have origins, such as lone-pair, geometric, charge ordering and of course spin driven mechanisms at least in perovskite compounds [10–13]. The second main problem is caused by low temperature of the phase transition associated with at least one of the ferroic orders thus hampering wide perspective of practical applications [4,14]. Additionally, some of multiferroic compounds are also characterized by quite weak interaction between magnetic and electric orders [15,16].

Widely known $BiFeO_3$ (BFO) perovskite could potentially assist in solving the aforementioned problems. The crystal structure of the compound is characterized by the rhombohedral symmetry (R3c space group) [17]. Additionally, $BiFeO_3$ possesses both, ferroelectric order at room temperature arising from the lone pair electrons of $Bi^{3+}$ ions and antiferromagnetic ordering caused by $Fe^{3+}$ ions. This makes bismuth ferrite a room temperature multiferroic compound with Neel temperature of 643 K and ferroelectric Currie temperature of around 1100 K [18–21]. However, there are a few significant drawbacks ascribed to this compound. Firstly, the preparation of single phase $BiFeO_3$ is quite complicated due to high volatility of bismuth as well as the prevalent formation of impurity phases such as $Bi_2Fe_4O_9$ and $Bi_{25}FeO_{40}$ [17]. Secondly, the compound is also characterized by large leakage current [22]. Lastly, the antiferromagnetic ordering is not as desirable as ferromagnetic one in multiferroic compound. To solve most of these issues, at least in part, chemical doping can be used. While there are numerous potential substitution options such as a formation of solid solutions as $BiFeO_3$-$BaTiO_3$, $BiFeO_3$-$NaNbO_3$ etc. or chemical doping with alkali earth metal elements (Ca, Mg), but the most common approach is doping with rare-earth (RE) (La-Lu) elements [23–27]. Doping of bismuth ferrite with RE elements has a couple of distinct advantages such as a stabilization of metastable structural state associated with an existence of polar active (R3c) to non-polar (Pnma) orthorhombic phases forming a morphotropic phase boundary (MPB) with an intermediate $PbZrO_3$-like anti-polar (Pbam) orthorhombic structure [17,28]. This leads to an exponential enhancement in the properties as well as increased sensitivity to external stimuli near the MPB [28,29]. Chemical doping with samarium ions is of particular interest as in this case the MPB is very narrow and the concentration range attributed to the single phase Pbam structure is reduced down to about 1 % [24,31]. Furthermore, the doping of BFO with rare-earth elements also leads to the disruption of the cycloidal antiferromagnetic ordering and causes a formation of weak ferromagnetic state due to the Dzyaloshinskii-Moriya interaction [32,33]. It is also known that post processing treatment with either high pressure or elevated temperatures or different synthesis procedures can also modify structural stability and in turn multiferroic properties of the compounds [34–39]. Overall, the ample variety of possible dopants as well as the increased sensitivity to external stimuli such as temperature, pressure, external electric and magnetic fields allows for a fine tuning of functional properties needed for the for the applications. However, the combined effect of the aforementioned modifications and the details of the temperature and pressure induced phase transitions are still needed to be further researched.

Hence in this work we provide an insight in the pressure induced phase transitions that occurred in the sol-gel derived samarium doped $BiFeO_3$ compounds with chemical compositions near the MPB. The structural comparison of the as prepared compounds as well as those subjected to high external pressure (P ~ 5 GPa) using different measurements techniques (laboratory and synchrotron X-ray diffraction, Raman and Mossbauer spectroscopy, SEM). The obtained data provide the detailed information about the changes in the crystal and magnetic structures of the $Bi_{1-x}Sm_xFeO_3$ compounds occurred under external pressure.

## 2. Experimental

The compounds were prepared using modified sol-gel synthesis method taking high purity $Bi(NO_3)_3 \cdot 5H_2O$, $Fe(NO_3)_3 \cdot 9H_2O$, $Sm_2O_3$ chemicals as starting materials, details of the synthesis procedure is described in our previous publications or can be found in supplementary material [24,27]. High pressure treatment (P ~ 5 GPa, duration time ~ 5 min) has been applied to the powder compounds after the synthesis, the samples were covered by molybdenum foil and graphite gaskets embedded in BN containers. Characterization of the samples was performed by means of X-ray diffraction (XRD) using a Rigaku MiniFlex II diffractometer with Ni filtered Cu Kα radiation in the 2theta range 10 - 70° with a step of 0.02°. The crystal structure of the compounds was also analyzed using synchrotron powder diffraction (SPD) measurements performed at KMC-2 beamline. Temperature dependent SPD data were recorded in the range 2Θ (10 - 100 °) with a step of 0.014 ° in the range of 20 – 700 °C. The diffraction data were analyzed by the Rietveld method using the FullProf software [40]. Scanning electron micrographs were taken with Hitachi SU-70 SEM and Tescan Vega 3 SEM. The particle size was measured using ImageJ software [41]. Raman spectra were recorded using inVia Raman spectrometer equipped with thermoelectrically cooled (-70 °C) CCD camera and microscope. Spectra were excited with 532 nm. To avoid damage of the sample, the laser power at the sample was restricted to very low value, 0.06 mW powder samples and 0.02 mW for samples affected by high pressure. The overall integration time was 800 s. Mössbauer spectrometer in transmission geometry using $^{57}Co(Rh)$ source was applied to collect spectra at room temperature. The Mössbauer spectra were fitted to sextet and singlet using WinNormos Site software. Isomer shifts are given relatively to α-Fe at room temperature. The pressure of 5 GPa was carried out at the installation, which includes a hydraulic press (DO 137A, Russia) and a high-pressure apparatus with two counter-moving carbide matrices of the 'anvil with a hole' type.

## 3. Results and Discussion
### 3.1. X-Ray diffraction analysis

The room temperature XRD data obtained for the as-prepared powder samples and the compounds (compacted powders) subjected to high pressure has confirmed the phase purity of the obtained compounds. Analysis of the diffraction patterns obtained for the powder samples and samples after pressing (Fig. 1) reveals distinct differences between the patterns of the both series pointing at strong effect of the high pressure on the crystal structure of the compounds.

The compounds containing 8 mol. % of samarium content in both series are characterized by a single phase structure with rhombohedral symmetry (R3c space group) which is characteristic for the initial bismuth ferrite [42]. Further chemical doping up to 12 mol. % of Sm content already induces slight changes in the crystal structure of the compounds in case of powder samples. Though the presence of other phases is not reliably detected by the diffraction measurements, a refinement of the diffraction pattern using two phases model assuming a coexistence of the rhombohedral (R3c) and anti-polar orthorhombic phase (Pbam) notably improves reliability factors and points at short-range character of the orthorhombic phase presented in the compound. The formation of an intermediate Pbam phase is distinctly confirmed by an appearance of the (110) and (112) reflections at around 17.7° and 28.8° (Fig. 1) when x = 0.14. Furthermore, the reflections characteristic to the rhombohedral phase (viz. (006), (113) at around 37.8° and 39.1°) drastically decrease in intensity while the intensity of the reflections ascribed to the orthorhombic phase increases, and already the orthorhombic phase comes to be dominant. The diffraction patterns of the compound with 16 mol. % of Sm show an appearance of the new reflections attributed to the non-polar orthorhombic phase described by the Pnma space group (reflections (111), (210) located at around 26.0° and 32.5°) which is typical structure for heavily doped $Bi_{1-x}Re_xFeO_3$ compounds [34]. Further increase in samarium content strengthens the trend and the volume fraction of the Pnma phase increases as compared to the Pbam phase and the compound with 20 mol. % of samarium content is characterized by single phase structure described by Pnma space group.

    In the case of pressed compounds, the transition to the orthorhombic phase occurs at lower level of Sm concentration. The main difference in the structure evolution observed for the compounds of the both series is related to a formation of the orthorhombic phases. In the compounds subjected to high pressure, the reflections specific for the anti-polar Pbam phase become notable already in the compound with 12 mol. % of samarium. At the same time, the amount of the Pbam phase is notably smaller in the compacted powder compounds having mixed structural state as compared to powder samples without high pressure treatment (Fig. 1). The non-polar orthorhombic phase is also stabilized in the HP compounds having lower Sm concentration as compared to the powder compounds thus pointing at more preferred conditions for the formation of the orthorhombic phases under high pressure as it provides a reduction in the unit cell parameters. This can be seen in particular in compounds at the edge of the MPB as for the powder containing 14 % of Sm the amount of Pbam phase is only ~83 % while in the case of compacted powder it is makes up ~92 %. Similar case was observed for the compound containing 18 % of Sm as in powder the amount of Pnma phase reaches about 70 % while for compacted powders it is about 95 %. Since different phases in the compounds around the phase boundary usually have relatively similar free energy the phase equilibrium can be affected by external stimuli. In this case in particular since high pressure was applied the equilibrium was shifted towards smaller volume unit cell containing phases.

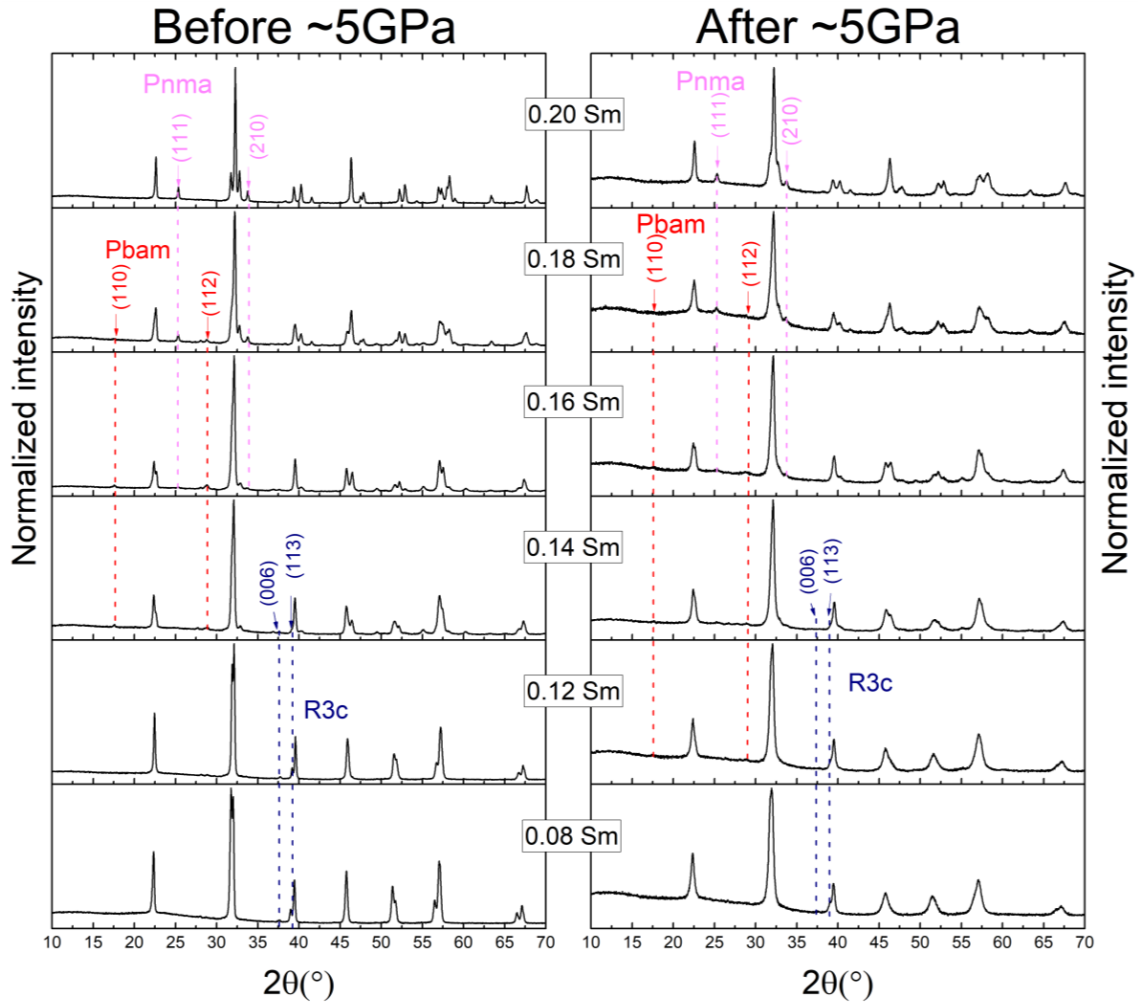

Fig.1. XRD patterns of $Bi_{1-x}Sm_xFeO_3$ compounds before and (left) after (right) application of high pressure.

In order to in detail study, the effects of both the chemical doping and high-pressure on the unit cell of the compounds, the structural parameters were calculated and analyzed. These results are presented in the Figure 2. Firstly, in both cases the substitution of $Bi^{3+}$ ions with $Sm^{3+}$ leads a contraction of the lattice which is caused by smaller ionic radius of the samarium ions [43,44]. The unit cell volumes calculated for the samples subjected to high pressure have slightly smaller values as compared to powder compounds. Based on the diffraction data, the concentration ranges of the different structural phase were determined and compared for the compounds of both series having the same chemical compositions. The concentration region attributed to the R3c structure lasts to the dopant level of 12 mol. %. The next concentration range attributed to the coexistence of the R3c and Pbam phases is determined in the region $0.12 \leq x \leq 0.15$. The third one is attributed to the coexistence of two different orthorhombic phases described by the space groups Pbam and Pnma - $0.15 < x < 0.20$. The single phase state with non-polar orthorhombic structure is attributed to the compound with x = 0.20. It is worthy to note that, the volume of the unit cell calculated to for the R3c phase is the highest while that one calculated for Pnma phase is the smallest. The mentioned evolution of the crystal structure did not reveal any evidence of extra structural phases while it points at important role of the anti-polar orthorhombic phase which serves as a bridge phase in the concentration and temperature driven phase transitions in BiFeO$_3$-based compounds [45].

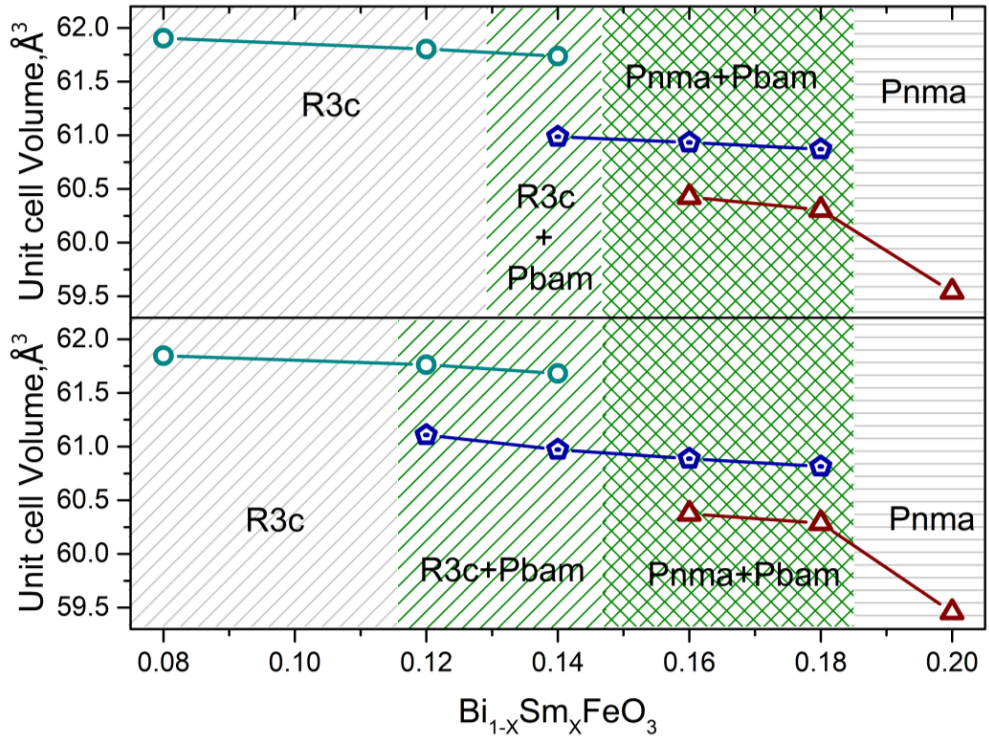

Fig.2. The unit cell volumes of the compounds Bi$_{1-x}$Sm$_x$FeO$_3$ at room temperature before (upper image) and after application of high pressure (unit cell volumes are presented in the reduced form, viz. $V_{R3c}^* = V_{R3c}/6$; $V_{Pbam}^* = V_{Pbam}/8$; $V_{Pnma}^* = V_{Pnma}/4$).

To provide further insight into the structural state of the compounds subjected to high pressure, the temperature dependent synchrotron diffraction measurements were performed for the compound with 12 mol. % of samarium which is characterized by a largest difference between powder and compacted powder compounds. For powder compound, even at temperature of ~300 K no reflections specific for Pbam phase were observed, indicating a single phase state with the rhombohedral symmetry described by R3c space group. The temperature range of the rhombohedral phase stability lasts up to 550 K when the formation of Pnma phase starts as confirmed by the appearance of the (111) reflection at around 25.0° (Figure 3). The temperature range ascribed to the mixed structural phase is rather narrow as at temperatures above 600 K the single-phase structure stabilizes. In the case of sample after application of high pressure, the presence of PbZrO$_3$-like Pbam phase is observed in much wider temperature range above 300 K. The reflections specific for this phase reach maximal intensity at temperature about 400 K and then start to decrease. Further increase in temperature leads to the formation of non-polar Pnma phase as confirmed by the appearance of the specific reflections indexed as (111) and (210) (Figure 3). At temperatures above 600 K there are no reflections specific for the Pbam phase and single phase Pnma state is stabilized.

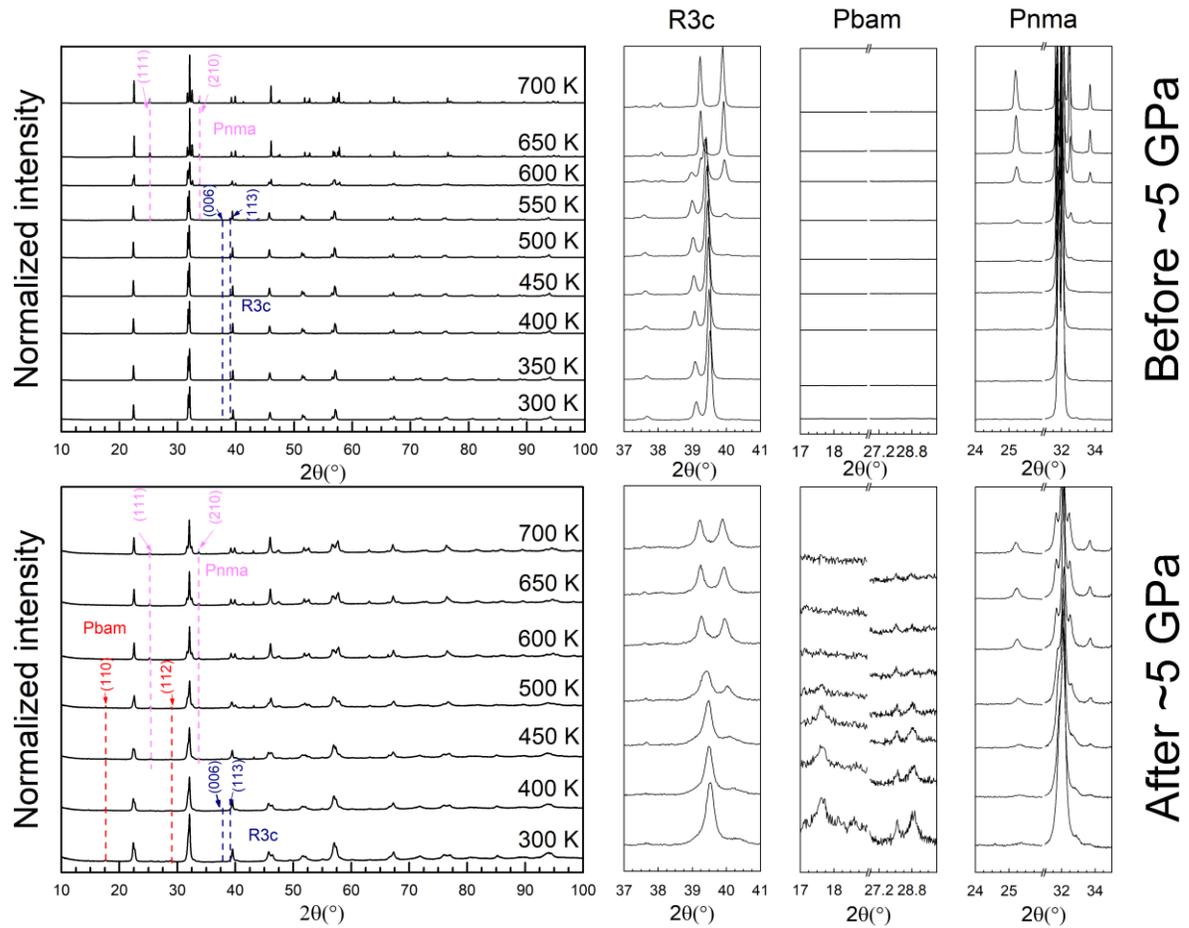

Fig.3. Synchrotron diffraction patterns of $Bi_{0.88}Sm_{0.12}FeO_3$ sample before and after application of pressure recorded at different temperatures. Insets denote the pattern where phase specific reflections are located. Left for R3c ((006) reflection) phase, middle for Pbam ((110) and (112) reflections) phase and right for Pnma ((111) and (210) reflections) phase.

The refined temperature dependent structural data (Figure 4) supports the aforementioned model of phase transitions. The obtained data confirm the existence of the different phase regions in both compacted and as prepared powders at various temperatures. In the case of powders, a by a single rhombohedral phase, which is stable up to a temperature of about 500 K. At higher temperatures the non-polar orthorhombic structure is stable at least up to 700 K. In the case of compacted powders, no single-phase rhombohedral structure is observed. Even at room temperature, a mixed structural state having both R3c and Pbam phases is determined. At 450 K a new mixed phase region is determined where the R3c phase is no longer exist and the phase described by Pnma space group is formed. Henceforth, a mixed phase regions with the non-polar and the anti-polar orthorhombic structures is observed in the temperature range from 450 to 550 K. The next phase region is determined to be of single phase with Pnma space group in the temperature range from 600 to 700 K. Furthermore, the thermal expansion of the unit cell is much higher for the R3c space group in the case of powders, while the other phases show similar thermal expansion.

Overall, the data obtained from synchrotron diffraction for the samples support the results obtained by conventional diffraction methods providing the additional information. The application of high pressure changes the stability of the phase mixture towards the crystal structures having smaller unit cell, this is especially noticeable by a stabilization of the PbZrO$_3$-like phase. Additionally, a major broadening of the diffraction peaks was also observed, which is most likely caused by the reduction in size of the particles after application of high pressure as confirmed by SEM measurements.

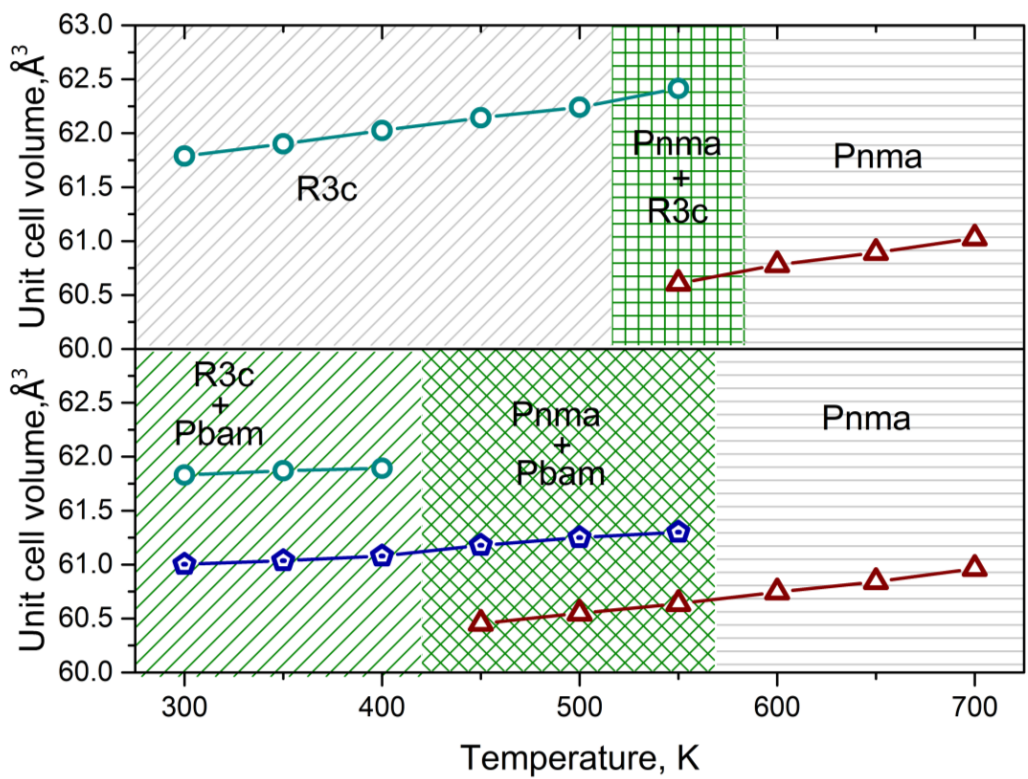

Fig.4. Dependency of unit cell volume on temperature for Bi$_{0.88}$Sm$_{0.12}$FeO$_3$ powder (top) and compacted powders (bottom) samples (unit cell volumes are presented in the reduced form, viz. V$_{R3c}$* = V$_{R3c}$/6;  V$_{Pbam}$* = V$_{Pbam}$/8; V$_{Pnma}$* = V$_{Pnma}$/4).

### 3.2. SEM data

To further investigate the effect of high pressure as well as the dopant concentration on the compounds the analysis using SEM technique was performed. This is especially important as particle size and morphology may potentially affect the phase stability of the compounds near the MPB [46,47]. Furthermore, magnetic and other properties needed for the for the applications of the compounds can also be influenced depending on the particle size and the morphology.

The SEM data obtained for both powder and compacted powder samples is displayed in the Figure 5. For the powder compound with 8 mol. % of samarium the particles are characterized by broad size distribution with the average size of around 0.77 µm. In the compound subjected to high pressure, the number of small particles is much larger reducing the average size down to about 0.65 µm. Increase in the samarium content up to 14 mol. % leads towards significant changes, viz. the average particle size decreases down to 0.50 µm for powder compounds. In the case of samples after application of high pressure, the presence of small particles also increases as compared to compounds having smaller samarium concentration and the average particle size is about 0.43 µm. For the compound with 20 mol. % of samarium, the average particle size remains similar, viz. about 0.50 µm for powders and about 0.46 µm for compacted powders. Furthermore, after the application of HP, the number small particles is much smaller as compared to that in the compound inside the MPB (e.g. 14 mol. % of Sm).

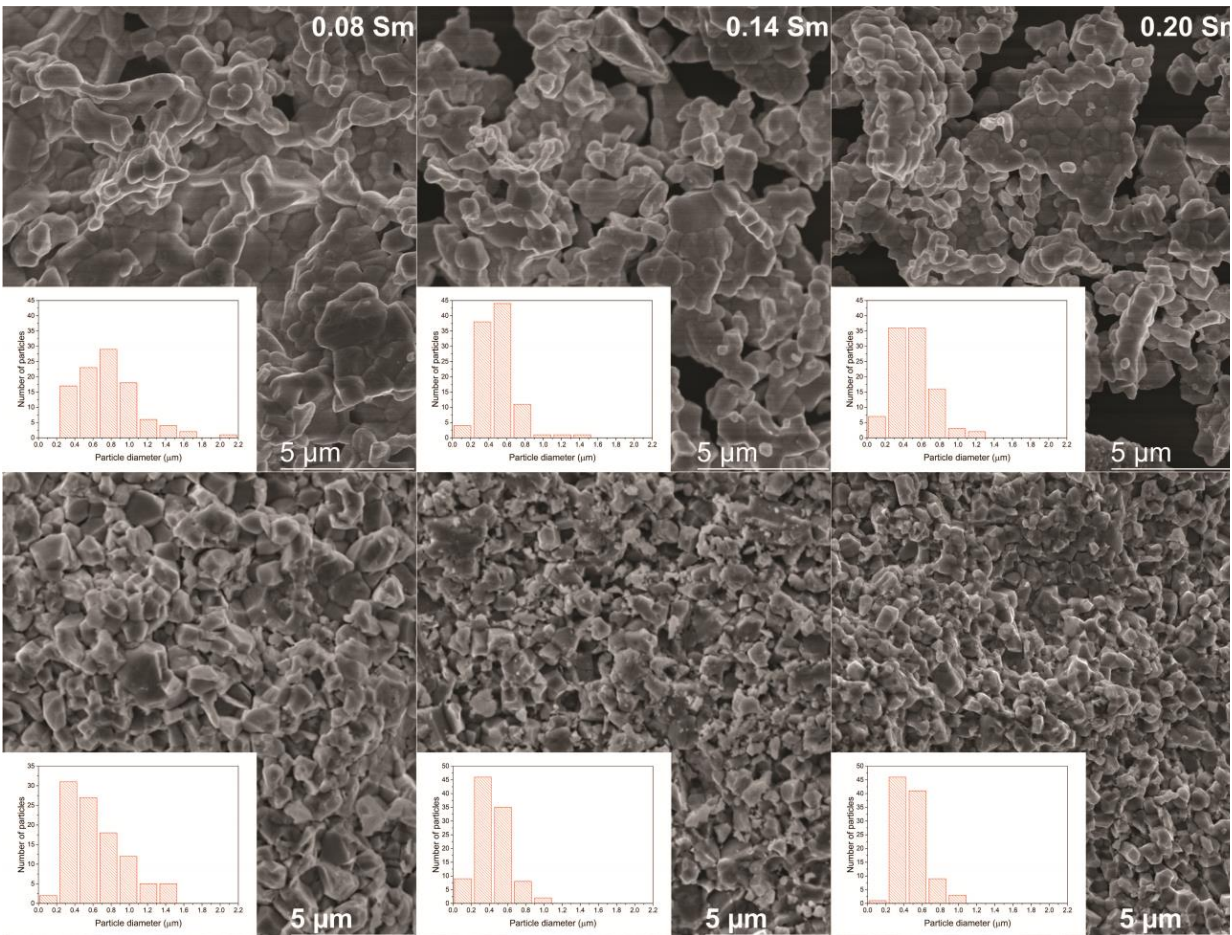

Fig.5. SEM images of powders (top row) and compacted powders (bottom row) of $Bi_{1-x}Sm_xFeO_3$ when x = 0.08, 0.14, 0.20. with particle size histogram insets.

Overall, the shape of the majority of the particles remains the same either for compacted or as prepared powders for all compounds and resembles the unit cell geometry (rectangular-like) of the structural phases [48]. The boundaries between the particles remain quite clear for all powder samples while in the case of samples after application of high pressure the borders become diffuse as the individual particles agglomerate into clusters, furthermore the particle size distribution is rather broad (from 0.05 µm up to 2.5 µm). However, large particle size distribution is quite common for compounds prepared by sol-gel synthesis route and sintered at elevated temperatures [49,50]. The change in particle size is most likely caused by either the fact that samarium ions are less chemically active as compared to bismuth ions thus reducing the mass transport speed resulting in smaller particles [51]. Another critical factor is high volatility of the bismuth ions which causes the formation of oxygen vacancies in bismuth rich compounds thus providing more intense grain growth and thus larger grains [52]. Overall, an application of high pressure provides a formation of smaller particles thus reducing the average particle size, which was most likely caused by the crushing of larger particles during the application of pressure, which is also confirmed by a broadening of the XRD reflections [53,54].

### 3.3. Raman spectroscopy results

Diffraction studies afford important but averaged data about the crystal structure of the studied compounds, while Raman spectroscopy provides structural information on local or short range level which is difficult to acquire by other structure-sensitive techniques [55]. In addition, Raman spectroscopy is very sensitive to presence of defects and disorders [23,49,56] and such information is of particular importance for the compounds having compositions near the phase boundaries. Figure 6 shows composition-induced changes in the Raman spectra of the initial compounds $Bi_{1-x}Sm_xFeO_3$ in the frequency region of 70–600 cm$^{-1}$. The well-defined bands characteristic for the rhombohedral structure of $Bi_{1-x}Sm_xFeO_3$ (x = 0.08) are visible at 78 cm$^{-1}$ (symmetry vibrational mode *E*), 145 cm$^{-1}$ (*A$_1$*), 176 cm$^{-1}$ (*A$_1$*), and 232 cm$^{-1}$ (*A$_1$*) [23,57]. The low intensity peaks at 474 and 527 cm$^{-1}$ belong to *A$_1$* and *E* symmetry vibrational modes, respectively [57]. First principles calculations predict vibrational modes associated with motion of mainly $Bi^{3+}$ ions at frequencies lower than 167 cm$^{-1}$, while the bands visible at frequencies higher than 262 cm$^{-1}$ are associated with oxygen ions [58]. It should be noted, that according to the diffraction data, the compound with 8 mol. % of samarium content has single-phase rhombohedral structure. Progressive upshift of these bands is notable for the compounds with higher dopant concentration. According to the XRD data, the compound with x = 0.12 is characterized by a single phase with polar rhombohedral (R3c) structure. The changes observed in the peak positions point at a perturbation in the crystal structure; however, the general spectral pattern remains similar, indicating that the rhombohedral structure is preserved. More notable structural perturbations are visible in the Raman spectrum for the compound with x = 0.14. The band at 178 cm$^{-1}$ became dominant in the studied spectral region; in addition, relative intensity of the bands located 284 and 530 cm$^{-1}$ increases (Figure 6). In this compound, relative content of the polar rhombohedral phase considerably decreases and the anti-polar orthorhombic becomes to be dominant confirming the results of the diffraction data. Further replacement of $Bi^{3+}$ ions by $Sm^{3+}$ up to 16% results in drastic changes in the crystal structure; viz. the low-frequency band near 150 cm$^{-1}$ characteristic for the rhombohedral structure disappears and strong band at 182 cm$^{-1}$ dominates in the spectrum. This compound possesses mixed structural state consisting the anti-polar (Pbam) and the non-polar (Pnma) orthorhombic phases; the rhombohedral (R3c) phase completely disappears. The XRD data confirm that further increase in Sm content results in a domination of the non-polar orthorhombic phase (x = 0.18 and 0.20). This new structure exhibits intense and broad Raman band near 303 cm$^{-1}$.

Furthermore, analysis of the Raman spectra allowed to recognize the specific Raman active bands characteristic to particular structural phase; estimate so-called Raman marker bands. These bands and corresponding tentative assignments are listed in Table S1 (supplementary). It is demonstrated that the frequency of so-called quasi-soft Raman bands in perovskites in the R3c and Pnma structures depends on the octahedral tilt angle [59,60], the frequencies increases with increasing the tilt angle. Importantly, the frequency upshift of these modes was observed immediately after addition of small amount of Sm$^{3+}$ ions. Thus, in addition to the XRD data the Raman spectroscopy data allowed to reveal very small perturbations in the symmetry of the polar rhombohedral phase on the local scale (see supplementary data).

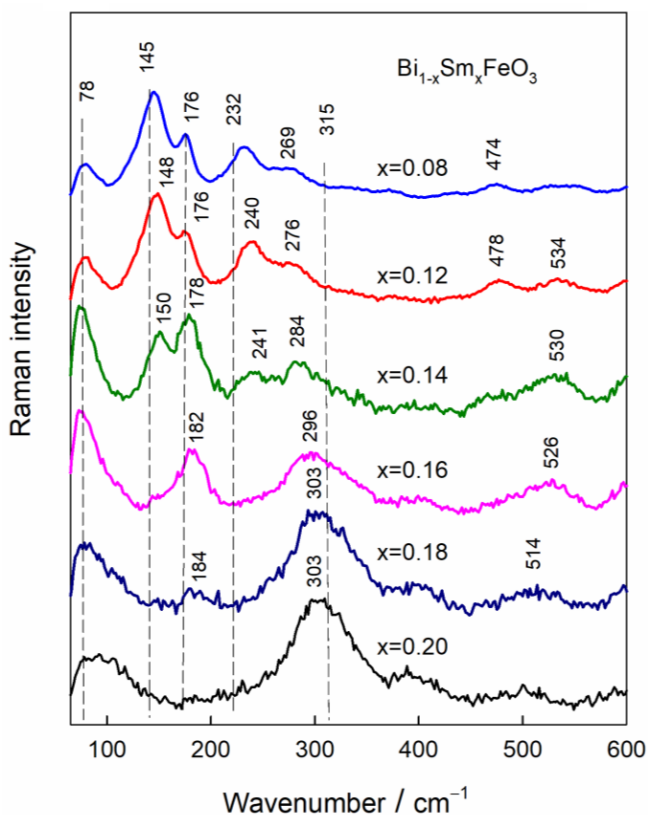

Fig.6. Composition dependent Raman spectra of polycrystalline Bi$_{1-x}$Sm$_x$FeO$_3$ compounds. Intensities are normalized to the intensity of the most intense band and spectra are shifted vertically for clarity. The excitation wavelength is 532 nm (0.06 mW).

Figure 7 compares composition-dependent Raman spectra of the compounds Bi$_{1-x}$Sm$_x$FeO$_3$ subjected to high pressure. The compounds with x = 0.08 and 0.12 retain their rhombohedral structure (R3c space group) as evident from the dominant characteristic band at 147–149 cm$^{-1}$ (Fig. 7, Table S1). However, important pressure-induced differences are visible in the Raman spectra for the compounds with x = 0.12, 0.14 and 0.18, comparing with the spectra of the initial compounds (Figure 6). The band characteristic for the anti-polar orthorhombic (Pbam) structure near 182–185 cm$^{-1}$ (x = 0.12, 0.14, 0.18) was detected in the spectra recorded for the pressure-affected samples as well as before the application of high pressure. This band was coupled with an appearance of the characteristic low intensity broad band at around 293 cm$^{-1}$, which is also specific for Pbam space group. However, in this case the peaks were much broader and more difficult to identify. Furthermore, a notable feature near 300 cm$^{-1}$ indicates the presence of the dominant non-polar orthorhombic phase (Table S1) already in the compound with x = 0.18. Thus, the Raman data confirm the structural model mentioned above and based on the X-ray diffraction data, that in the compounds subjected to high pressure the intermediate non-polar orthorhombic and anti-polar phases form at lower concentration of the dopant ions.

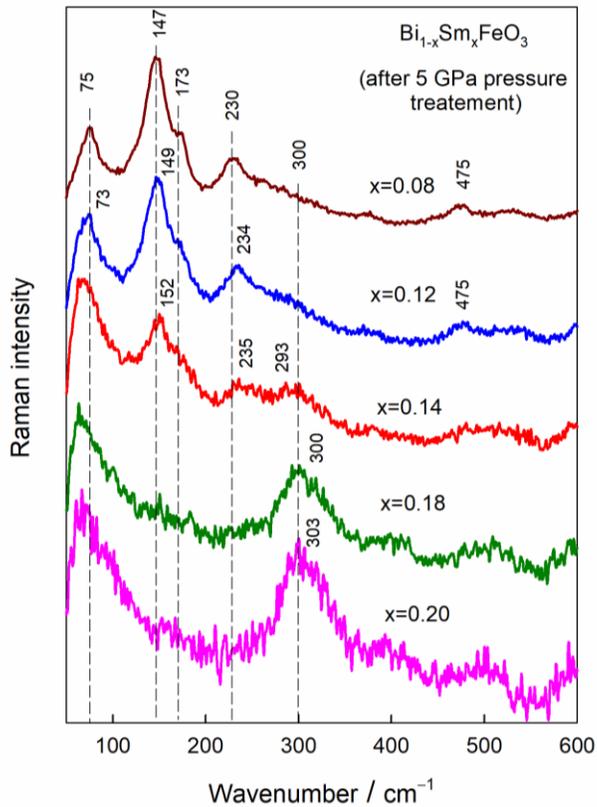

Fig.7. Composition-dependent Raman spectra of polycrystalline Bi$_{1-x}$Sm$_x$FeO$_3$ compounds affected by high pressure. Intensities are normalized to the intensity of the most intense band and spectra are shifted vertically for clarity. The excitation wavelength is 532 nm (0.02 mW).

### 3.4. Mössbauer spectroscopy results

Mössbauer spectra, measured at room temperature, of the two series of samples before and after application of high pressure are shown in the Figure 8. Spectral parameters are presented in Tables S2 and S3. Mössbauer spectra are fitted to one sextet and one broad singlet. The majority of spectral area belonged to sextet which had isomer shift $\delta$=0.39±0.01 mm/s, hyperfine field $B$=49.4-49.96 T, linewidth $\Gamma$= 0.3-0.55 mm/s and quadrupole shift $2\varepsilon$=- 0.1-0.05 mm/s. The variations of hyperfine field and quadrupole shift can be caused by the changes in crystalline structure of $Bi_{1-x}Sm_xFeO_3$ and local symmetry of Fe atom neighborhood, which is disturbed when Bi is substituted by Sm.

In magnetically ordered phase the quadrupole interactions cause small shifts $\varepsilon$ of sextet lines. Quadrupole shift $2\varepsilon = \frac{eQV_{zz}}{4}(3\cos^2\theta - 1 + \eta\sin^2\theta\cos2\varphi)$ [61] depends on the main component of electric field gradient (EFG) $V_{zz}$, nuclear quadrupole moment Q, asymmetry parameter of EFG, and the angles $\theta$ and $\phi$. The angle $\theta$ is the angle between magnetization direction and EFG z axis while the angle $\phi$ is between magnetization projection into the xy plane and EFG x axis. The amplitudes of components of EFG increase with the decrease of crystalline symmetry while $V_{zz} + V_{xx} + V_{yy} = 0$ and the EFG axis are chosen so that $|V_{zz}| \geq |V_{xx}| \geq |V_{yy}|$. It can be noted, that quadrupole shift $2\varepsilon$ depends not only on quadrupole interaction parameter but also on the direction of magnetization with respect to the EFG axis.

It should be noted, that for initial $BiFeO_3$ the sextet spectrum is asymmetrically broadened because of spiral magnetic structure and related change of Fe spin orientation [62]. The Mössbauer spectrum of $BiFeO_3$ is fitted well using two sextets having different hyperfine fields and quadrupole shifts B=49.0; 50.3 T and $2\varepsilon$=0.02 and 0.22 mm/s [63]. Average values for $BiFeO_3$: B=49.7 T and $2\varepsilon$=0.13 mm/s indicate similar hyperfine field and larger quadrupole shift in comparison with $Bi_{1-x}Sm_xFeO_3$ samples.

The spectral asymmetry was not characteristic of $Bi_{1-x}Sm_xFeO_3$ Mössbauer spectra (Figure 8) indicating that dominant crystalline and/or magnetic structure is different than that of $BiFeO_3$. The large linewidth for 8 % mol Sm sample (Tables S2 and S3) indicate that similar variations of the hyperfine field and quadrupole shift exist as in pure $BiFeO_3$ but the spectral asymmetry disappears due to a different interrelation of the parameters and a different magnetic order. The increase in linewidth which is more characteristic for samples with lower Sm content can be caused by distortions of local Fe neighborhood symmetry when Bi is substituted by Sm what leads to changes in quadrupole shift. Moreover, structural transition to the orthorhombic phase as denoted by diffraction data causes modification in the spectral contributions characterized by different quadrupole shifts and hyperfine fields which cause spectral broadening. However, linewidth $\Gamma$ for 20 % mol of Sm sample was close to characteristic linewidth of α-Fe calibration foil ($\approx$ 0.3 mm/s) which indicated fairly similar local environment of Fe and the same crystalline structure.

For the samples subjected to high pressure, the value of quadrupole shift is negative and more constant with change in Sm percentage indicating different local environment of Fe and the crystal structure comparing with initial powder samples. However, smaller linewidth and the decrease in absolute values of quadrupole shift for the high pressure compounds with samarium content more than 14 mol. % may be associated with the increase of contribution of the non-polar orthorhombic structure which is in accordance with the results of the diffraction measurements and the magnetization data [24]. It is noteworthy that when using one sextet for spectrum fitting the sextets parameters (quadrupole shift, hyperfine field and isomer shift) would be averaged due to contributions of different structures (R3c, Pbam, Pnma) which may like R3c have different parameters at different local positions. The sextet lines of the compacted powder compounds are broader and spectral contribution of paramagnetic singlet is larger thus denoting more pronounced amount of the structural fraction having poor crystallization. The structural fraction having lower crystallization is caused by extended amount of the upper surface layer of the crystallites as compared to the morphology of the initial powder compounds.

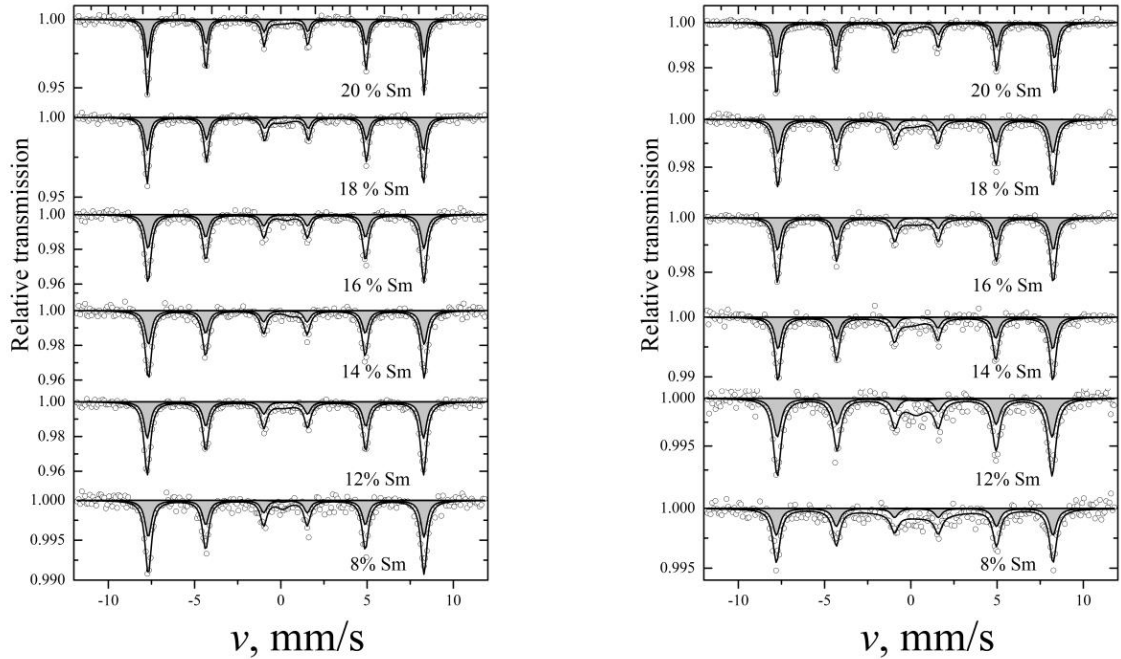

Fig. 8. Mössbauer spectra of BiFeO$_3$ a with Sm before high pressure (left) and after high pressure (right).

## Conclusions

The obtained results testify that an application of external pressure to the Sm-doped BiFeO$_3$ compacted powders with compositions in the morphotropic phase boundary provides a stabilization of the non-polar orthorhombic phase. The amount of the anti-polar orthorhombic phase considered as intermediate phase which bridges the structural transition from the rhombohedral to the non-polar orthorhombic phase notably decreases wherein the local scale measurements specify its presence across the concentration driven phase transition. Temperature increase causes the structural transition to the nonpolar orthorhombic phase in the studied compounds regardless their structural state stable at room temperature, wherein an application of external pressure decreases the temperature of the mentioned phase transition. Mössbauer spectroscopy data has determined a modification of the local symmetry of iron ions associated with orthorhombic distortion and related changes in the magnetic state of the compounds subjected to external pressure; an increase in the amount of paramagnetic fraction with the dopant content is associated with enlarged amount of grain surface having poor crystallinity in the pressed compounds.

## Acknowledgments

This project has received funding from the European Union's Horizon 2020 research and innovation programme under the Marie Skłodowska-Curie grant agreement No 778070 – TransFerr – H2020-MSCA-RISE-2017. G.N. gratefully acknowledges the Center of Spectroscopic Characterization of Materials and Electronic/Molecular Processes (SPECTROVERSUM Infrastructure) for use of Raman spectrometer. A.V.S. and D.V.K. acknowledge RFBR (projects # 20-58-00030) and BRFFR (project # F20R-123). S.I.L. and V.V.S. acknowledge RFBR (projects # 20-52-00023) and BRFFR (project # T20R-121). The authors acknowledge Helmholtz-Zentrum Berlin für Materialien und Energie for the allocation of synchrotron radiation beamtime and HZB staff for the assistance with synchrotron radiation experiments.

## References:

[1] I.H. Lone, J. Aslam, N.R.E. Radwan, A.H. Bashal, A.F.A. Ajlouni, A. Akhter, Multiferroic ABO3 Transition Metal Oxides: a Rare Interaction of Ferroelectricity and Magnetism, Nanoscale Res. Lett. 14 (2019) 1–12. https://doi.org/10.1186/s11671-019-2961-7.

[2] H. Hou, P. Finkel, M. Staruch, J. Cui, I. Takeuchi, Ultra-low-field magneto-elastocaloric cooling in a multiferroic composite device, Nat. Commun. 9 (2018) 1–8. https://doi.org/10.1038/s41467-018-06626-y.

[3] M. Kumar, A. Kumar, A. Anshul, S. Sharma, Advances and future challenges in multifunctional nanostructures for their role in fast, energy efficient memory devices, Mater. Lett. 277 (2020) 128369. https://doi.org/10.1016/j.matlet.2020.128369.

[4] N.A. Spaldin, R. Ramesh, Advances in magnetoelectric multiferroics, Nat. Mater. 18 (2019) 203–212. https://doi.org/10.1038/s41563-018-0275-2.

[5] T. Lottermoser, T. Lonkai, U. Amann, D. Hohlwein, J. Ihringer, M. Fiebig, Magnetic phase control by an electric field, Nature. 430 (2004) 541–544. https://doi.org/10.1038/nature02728.

[6] P. Ravindran, R. Vidya, A. Kjekshus, H. Fjellvåg, O. Eriksson, Theoretical investigation of magnetoelectric behavior in BiFe O3, Phys. Rev. B - Condens. Matter Mater. Phys. 74 (2006) 224412. https://doi.org/10.1103/PhysRevB.74.224412.

[7] W. Eerenstein, N.D. Mathur, J.F. Scott, Multiferroic and magnetoelectric materials, Nature. 442 (2006) 759–765. https://doi.org/10.1038/nature05023.

[8] N.A. Spaldin, S.W. Cheong, R. Ramesh, Multiferroics: Past, present, and future, Phys. Today. 63 (2010) 38–43. https://doi.org/10.1063/1.3502547.

[9] N.A.N. Hill, Why are there so few magnetic ferroelectrics?, 104 (2000) 6694–6709. https://doi.org/10.1021/jp000114x.

[10] M. Fiebig, T. Lottermoser, D. Meier, M. Trassin, The evolution of multiferroics, Nat. Rev. Mater. 1 (2016) 1–14. https://doi.org/10.1038/natrevmats.2016.46.

[11] E.H. Smith, N.A. Benedek, C.J. Fennie, Interplay of Octahedral Rotations and Lone Pair Ferroelectricity in CsPbF3, Inorg. Chem. 54 (2015) 8536–8543. https://doi.org/10.1021/acs.inorgchem.5b01213.

[12] A.T. Mulder, N.A. Benedek, J.M. Rondinelli, C.J. Fennie, Turning ABO3 antiferroelectrics into ferroelectrics: Design rules for practical rotation-driven ferroelectricity in double perovskites and A3B2O7 Ruddlesden-popper compounds, Adv. Funct. Mater. 23 (2013) 4810–4820. https://doi.org/10.1002/adfm.201300210.

[13]     T. Tohei, H. Moriwake, H. Murata, A. Kuwabara, R. Hashimoto, T. Yamamoto, I. Tanaka, Geometric ferroelectricity in rare-earth compounds R GaO3 and R InO3, Phys. Rev. B - Condens. Matter Mater. Phys. 79 (2009) 144125. https://doi.org/10.1103/PhysRevB.79.144125.

[14]     H. Liu, X. Yang, A brief review on perovskite multiferroics, Ferroelectrics. 507 (2017). https://doi.org/10.1080/00150193.2017.1283171.

[15]     S. Dong, H. Xiang, E. Dagotto, Magnetoelectricity in multiferroics: A theoretical perspective, Natl. Sci. Rev. 6 (2019) 629–641. https://doi.org/10.1093/nsr/nwz023.

[16]     M. Li, H. Tan, W. Duan, Hexagonal rare-earth manganites and ferrites: A review of improper ferroelectricity, magnetoelectric coupling, and unusual domain walls, Phys. Chem. Chem. Phys. 22 (2020) 14415–14432. https://doi.org/10.1039/d0cp02195d.

[17]     T. Rojac, A. Bencan, B. Malic, G. Tutuncu, J.L. Jones, J.E. Daniels, D. Damjanovic, $BiFeO_3$ Ceramics: Processing, Electrical, and Electromechanical Properties, J. Am. Ceram. Soc. 97 (2014) 1993–2011. https://doi.org/10.1111/jace.12982.

[18]     V.R. Reddy, D. Kothari, A. Gupta, S.M. Gupta, Study of weak ferromagnetism in polycrystalline multiferroic Eu doped bismuth ferrite, Appl. Phys. Lett. 94 (2009) 82505. https://doi.org/10.1063/1.3089577.

[19]     I. Sosnowska, M. Loewenhaupt, W.I.F. David, R.M. Ibberson, Investigation of the unusual magnetic spiral arrangement in BiFeO3, Phys. B Phys. Condens. Matter. 180–181 (1992) 117–118. https://doi.org/10.1016/0921-4526(92)90678-L.

[20]     Z. Yu, J. Zeng, L. Zheng, W. Liu, G. Li, A. Kassiba, Large piezoelectricity and high Curie temperature in novel bismuth ferrite-based ferroelectric ceramics, J. Am. Ceram. Soc. 103 (2020) 6435–6444. https://doi.org/10.1111/jace.17382.

[21]     J. Walker, H. Simons, D.O. Alikin, A.P. Turygin, V.Y. Shur, A.L. Kholkin, H. Ursic, A. Bencan, B. Malic, V. Nagarajan, T. Rojac, Dual strain mechanisms in a lead-free morphotropic phase boundary ferroelectric, Sci. Rep. 6 (2016) 1–8. https://doi.org/10.1038/srep19630.

[22]     W. Xing, M. Yinina, Z. Ma, Y. Bai, J. Chen, S. Zhao, Improved ferroelectric and leakage current properties of Er-doped BiFeO3 thin films derived from structural transformation, Smart Mater. Struct. 23 (2014) 85030. https://doi.org/10.1088/0964-1726/23/8/085030.

[23]     A. Pakalniškis, A. Lukowiak, G. Niaura, P. Głuchowski, D.V. Karpinsky, D.O. Alikin, A.S. Abramov, A. Zhaludkevich, M. Silibin, A.L. Kholkin, R. Skaudžius, W. Strek, A. Kareiva, Nanoscale ferroelectricity in pseudo-cubic sol-gel derived barium titanate - bismuth ferrite (BaTiO3– BiFeO3) solid solutions, J. Alloys Compd. (2020) 154632. https://doi.org/10.1016/j.jallcom.2020.154632.

[24]     D. V. Karpinsky, A. Pakalniškis, G. Niaura, D. V. Zhaludkevich, A.L. Zhaludkevich, S.I. Latushka, M. Silibin, M. Serdechnova, V.M. Garamus, A. Lukowiak, W. Stręk, M. Kaya, R. Skaudžius, A. Kareiva, Evolution of the crystal structure and magnetic properties of Sm-doped BiFeO3 ceramics across the phase boundary region, Ceram. Int. (2020). https://doi.org/10.1016/j.ceramint.2020.10.120.

[25]     L. V. Costa, M.G. Ranieri, M. Cilense, E. Longo, A.Z. Simões, Evidence of magnetoelectric coupling on calcium doped bismuth ferrite thin films grown by chemical solution deposition, in: J. Appl. Phys., American Institute of Physics Inc., 2014: p. 17D910. https://doi.org/10.1063/1.4867123.

[26]     I.P. Raevski, S.P. Kubrin, J.L. Dellis, S.I. Raevskaya, D.A. Sarychev, V.G. Smotrakov, V. V. Eremkin, M.A. Seredkina, Studies of magnetic and ferroelectric phase transitions in BiFeO 3-NaNbO3 solid solution ceramics, in: Ferroelectrics, 2008: pp. 113–118. https://doi.org/10.1080/00150190802397767.

[27]     A. Pakalniškis, R. Skaudžius, D. V. Zhaludkevich, A.L. Zhaludkevich, D.O. Alikin, A.S. Abramov, T. Murauskas, V.Y. Shur, A.A. Dronov, M. V. Silibin, A. Selskis, R. Ramanauskas, A. Lukowiak, W. Strek, D. V. Karpinsky, A. Kareiva, Morphotropic phase boundary in Sm-substituted BiFeO3 ceramics: Local vs microscopic approaches, J. Alloys Compd. 875 (2021) 159994. https://doi.org/10.1016/j.jallcom.2021.159994.

[28]     D. V. Karpinsky, I.O. Troyanchuk, M. Tovar, V. Sikolenko, V. Efimov, A.L. Kholkin, Evolution of crystal structure and

ferroic properties of La-doped BiFeO3 ceramics near the rhombohedral-orthorhombic phase boundary, J. Alloys Compd. 555 (2013) 101–107. https://doi.org/10.1016/j.jallcom.2012.12.055.

[29]  J. Rödel, J.F. Li, Lead-free piezoceramics: Status and perspectives, MRS Bull. 43 (2018) 576–580. https://doi.org/10.1557/mrs.2018.181.

[30]  D. Damjanovic, A morphotropic phase boundary system based on polarization rotation and polarization extension, Appl. Phys. Lett. 97 (2010). https://doi.org/10.1063/1.3479479.

[31]  M. Kubota, K. Oka, Y. Nakamura, H. Yabuta, K. Miura, Y. Shimakawa, M. Azuma, Sequential Phase Transitions in Sm Substituted BiFeO 3 , Jpn. J. Appl. Phys. 50 (2011) 09NE08. https://doi.org/10.7567/jjap.50.09ne08.

[32]  C. Ederer, N.A. Spaldin, Weak ferromagnetism and magnetoelectric coupling in bismuth ferrite, Phys. Rev. B - Condens. Matter Mater. Phys. 71 (2005) 060401. https://doi.org/10.1103/PhysRevB.71.060401.

[33]  A.S. Erchidi Elyacoubi, R. Masrour, A. Jabar, Magnetocaloric effect and magnetic properties in SmFe1-xMnxO3 perovskite: Monte Carlo simulations, Solid State Commun. 271 (2018) 39–43. https://doi.org/10.1016/j.ssc.2017.12.015.

[34]  I.O. Troyanchuk, D. V. Karpinsky, M. V. Bushinsky, O.S. Mantytskaya, N. V. Tereshko, V.N. Shut, Phase transitions, magnetic and piezoelectric properties of rare-earth-substituted BiFeO3ceramics, J. Am. Ceram. Soc. 94 (2011) 4502–4506. https://doi.org/10.1111/j.1551-2916.2011.04780.x.

[35]  D.O. Alikin, A.P. Turygin, J. Walker, T. Rojac, V. V. Shvartsman, V.Y. Shur, A.L. Kholkin, Quantitative phase separation in multiferroic Bi0.88Sm0.12FeO3 ceramics via piezoresponse force microscopy, J. Appl. Phys. 118 (2015) 072004. https://doi.org/10.1063/1.4927812.

[36]  I.I. Makoed, A.A. Amirov, N.A. Liedienov, A. V. Pashchenko, K.I. Yanushkevich, D. V. Yakimchuk, E.Y. Kaniukov, Evolution of structure and magnetic properties in EuxBi1–xFeO3 multiferroics obtained under high pressure, J. Magn. Magn. Mater. 489 (2019) 165379. https://doi.org/10.1016/J.JMMM.2019.165379.

[37]  N.A. Liedienov, A. V. Pashchenko, V.A. Turchenko, V.Y. Sycheva, A. V. Voznyak, V.P. Kladko, A.I. Gudimenko, D.D. Tatarchuk, Y. V. Didenko, I. V. Fesych, I.I. Makoed, A.T. Kozakov, G.G. Levchenko, Liquid-phase sintered bismuth ferrite multiferroics and their giant dielectric constant, Ceram. Int. 45 (2019) 14873–14879. https://doi.org/10.1016/J.CERAMINT.2019.04.220.

[38]  A. V. Pashchenko, N.A. Liedienov, Q. Li, I.I. Makoed, D.D. Tatarchuk, Y. V. Didenko, A.I. Gudimenko, V.P. Kladko, L. Jiang, L. Li, V.G. Pogrebnyak, G.G. Levchenko, Control of dielectric properties in bismuth ferrite multiferroic by compacting pressure, Mater. Chem. Phys. 258 (2021) 123925. https://doi.org/10.1016/J.MATCHEMPHYS.2020.123925.

[39]  A. V. Pashchenko, N.A. Liedienov, Q. Li, D.D. Tatarchuk, V.A. Turchenko, I.I. Makoed, V.Y. Sycheva, A. V. Voznyak, V.P. Kladko, A.I. Gudimenko, Y. V. Didenko, A.T. Kozakov, G.G. Levchenko, Structure, non-stoichiometry, valence of ions, dielectric and magnetic properties of single-phase Bi0.9La0.1FeO3−δ multiferroics, J. Magn. Magn. Mater. 483 (2019) 100–113. https://doi.org/10.1016/J.JMMM.2019.03.095.

[40]  J. Rodríguez-Carvajal, Recent advances in magnetic structure determination by neutron powder diffraction, Phys. B Phys. Condens. Matter. 192 (1993) 55–69. https://doi.org/10.1016/0921-4526(93)90108-I.

[41]  J. Schindelin, I. Arganda-Carreras, E. Frise, V. Kaynig, M. Longair, T. Pietzsch, S. Preibisch, C. Rueden, S. Saalfeld, B. Schmid, J.Y. Tinevez, D.J. White, V. Hartenstein, K. Eliceiri, P. Tomancak, A. Cardona, Fiji: An open-source platform for biological-image analysis, Nat. Methods. 9 (2012) 676–682. https://doi.org/10.1038/nmeth.2019.

[42]  D. Arnold, Composition-driven structural phase transitions in rare-earth-doped bifeo3 ceramics: A review, IEEE Trans. Ultrason. Ferroelectr. Freq. Control. 62 (2015) 62–82. https://doi.org/10.1109/TUFFC.2014.006668.

[43]  Y.R. Sun, X. Zhang, L.G. Wang, Z.K. Liu, N. Kang, N. Zhou, W.L. You, J. Li, X.F. Yu, Lattice contraction tailoring in perovskite oxides towards improvement of oxygen electrode catalytic activity, Chem. Eng. J. 421 (2021) 129698. https://doi.org/10.1016/J.CEJ.2021.129698.

[44]     Y.Q. Jia, Crystal radii and effective ionic radii of the rare earth ions, J. Solid State Chem. 95 (1991) 184–187. https://doi.org/10.1016/0022-4596(91)90388-X.

[45]     M. Ahart, M. Somayazulu, R.E. Cohen, P. Ganesh, P. Dera, H. Mao, R.J. Hemley, Y. Ren, P. Liermann, Z. Wu, Origin of morphotropic phase boundaries in ferroelectrics., Nature. 451 (2008) 545–8. https://doi.org/10.1038/nature06459.

[46]     W. Jo, T.H. Kim, D.Y. Kim, S.K. Pabi, Effects of grain size on the dielectric properties of Pb (Mg1/3 Nb2/3) O3 -30 mol % PbTiO3 ceramics, J. Appl. Phys. 102 (2007) 074116. https://doi.org/10.1063/1.2794377.

[47]     C. Behera, R.N.P. Choudhary, P.R. Das, Structural, electrical and multiferroic characteristics of thermo-mechanically fabricated BiFeO3-(BaSr)TiO3 solid solutions, Mater. Res. Express. 5 (2018) 056301. https://doi.org/10.1088/2053-1591/aabeae.

[48]     M.K. Singh, First principle study of crystal growth morphology: An application to crystalline urea, (2006). http://arxiv.org/abs/cond-mat/0602385 (accessed July 29, 2020).

[49]     D. Karoblis, A. Zarkov, K. Mazeika, D. Baltrunas, G. Niaura, A. Beganskiene, A. Kareiva, Sol-gel synthesis, structural, morphological and magnetic properties of BaTiO3–BiMnO3 solid solutions, Ceram. Int. 46 (2020) 16459–16464. https://doi.org/10.1016/j.ceramint.2020.03.209.

[50]     G. Inkrataite, M. Kemere, A. Sarakovskis, R. Skaudzius, Influence of boron on the essential properties for new generation scintillators, J. Alloys Compd. 875 (2021) 160002. https://doi.org/10.1016/j.jallcom.2021.160002.

[51]     S. Cho, C. Yun, Y.S. Kim, H. Wang, J. Jian, W. Zhang, J. Huang, X. Wang, H. Wang, J.L. MacManus-Driscoll, Strongly enhanced dielectric and energy storage properties in lead-free perovskite titanate thin films by alloying, Nano Energy. 45 (2018) 398–406. https://doi.org/10.1016/j.nanoen.2018.01.003.

[52]     S. Liu, H. Luo, S. Yan, L. Yao, J. He, Y. Li, L. He, S. Huang, L. Deng, Effect of Nd-doping on structure and microwave electromagnetic properties of BiFeO3, J. Magn. Magn. Mater. 426 (2017) 267–272. https://doi.org/10.1016/j.jmmm.2016.11.080.

[53]     A.A. Al-Tabbakh, N. Karatepe, A.B. Al-Zubaidi, A. Benchaabane, N.B. Mahmood, Crystallite size and lattice strain of lithiated spinel material for rechargeable battery by X-ray diffraction peak-broadening analysis, Int. J. Energy Res. 43 (2019) 1903–1911. https://doi.org/10.1002/er.4390.

[54]     R. Yogamalar, R. Srinivasan, A. Vinu, K. Ariga, A.C. Bose, X-ray peak broadening analysis in ZnO nanoparticles, Solid State Commun. 149 (2009) 1919–1923. https://doi.org/10.1016/j.ssc.2009.07.043.

[55]     S. Gupta, R. Medwal, S.P. Pavunny, D. Sanchez, R.S. Katiyar, Temperature dependent Raman scattering and electronic transitions in rare earth SmFeO3, Ceram. Int. 44 (2018) 4198–4203. https://doi.org/10.1016/j.ceramint.2017.11.223.

[56]     R. Trusovas, G. Račiukaitis, G. Niaura, J. Barkauskas, G. Valušis, R. Pauliukaite, Recent Advances in Laser Utilization in the Chemical Modification of Graphene Oxide and Its Applications, Adv. Opt. Mater. 4 (2016) 37–65. https://doi.org/10.1002/adom.201500469.

[57]     A. Ahlawat, S. Satapathy, V.G. Sathe, R.J. Choudhary, M.K. Singh, R. Kumar, T.K. Sharma, P.K. Gupta, Modification in structure of La and Nd co-doped epitaxial BiFeO $_3$ thin films probed by micro Raman spectroscopy, J. Raman Spectrosc. 46 (2015) 636–643. https://doi.org/10.1002/jrs.4701.

[58]     P. Hermet, M. Goffinet, J. Kreisel, P. Ghosez, Raman and infrared spectra of multiferroic bismuth ferrite from first principles, 75 (2007). https://doi.org/10.1103/PhysRevB.75.220102.

[59]     N.D. Todorov, M. V. Abrashev, V.G. Ivanov, Frequency dependence of the quasi-soft Raman-active modes in rotationally distorted R 3+B 3+O 3 perovskites (R 3+rare earth, B 3+=Al, Sc, Ti, V, Cr, Mn, Fe, Co, Ni, Ga), J. Phys. Condens. Matter. 24 (2012) 8. https://doi.org/10.1088/0953-8984/24/17/175404.

[60]     J. Bielecki, P. Svedlindh, D.T. Tibebu, S. Cai, S.G. Eriksson, L. Börjesson, C.S. Knee, Structural and magnetic properties of isovalently substituted multiferroic BiFeO 3: Insights from Raman spectroscopy, Phys. Rev. B - Condens. Matter Mater. Phys. 86 (2012) 184422. https://doi.org/10.1103/PhysRevB.86.184422.

[61]     Y.L. Chen, D.P. Yang, Mössbauer Effect in Lattice Dynamics: Experimental Techniques and Applications, Mössbauer Eff. Lattice Dyn. Exp. Tech. Appl. (2007) 1–409. https://doi.org/10.1002/9783527611423.

[62]     D. Lebeugle, D. Colson, A. Forget, M. Viret, P. Bonville, J.F. Marucco, S. Fusil, Room-temperature coexistence of large electric polarization and magnetic order in <span class, Phys. Rev. B. 76 (2007) 024116. https://doi.org/10.1103/PhysRevB.76.024116.

[63]     D. Karoblis, D. Griesiute, K. Mazeika, D. Baltrunas, D. V. Karpinsky, A. Lukowiak, P. Gluchowski, R. Raudonis, A. Katelnikovas, A. Zarkov, A. Kareiva, A Facile Synthesis and Characterization of Highly Crystalline Submicro-Sized BiFeO3, Mater. 2020, Vol. 13, Page 3035. 13 (2020) 3035. https://doi.org/10.3390/MA13133035.